# Ultrafast Tracking of Exciton and Charge Carrier Transport in Optoelectronic Materials on the Nanometer Scale

*Christoph Schnedermann[†,a,\*], Jooyoung Sung[†,a], Raj Pandya[†], Sachin Dev Verma[†], Richard Y. S. Chen[†], Nicolas Gauriot[†], Hope Bretscher[†], Philipp Kukura[‡], Akshay Rao[†,\*]*

[†]Department of Physics, Cavendish Laboratory, University of Cambridge, JJ Thompson Avenue, Cambridge, CB3 0HE, United Kingdom

[‡]Physical and Theoretical Chemistry Laboratory, University of Oxford, South Parks Road, Oxford OX1 3QZ, United Kingdom

AUTHOR INFORMATION

[a] These authors contributed equally to the work.

**Corresponding Authors**

Christoph Schnedermann:     cs2002@cam.ac.uk

Akshay Rao:     ar525@cam.ac.uk

ABSTRACT

We present a novel optical transient absorption and reflection microscope based on a diffraction-limited pump pulse in combination with a wide-field probe pulse, for the spatio-temporal investigation of ultrafast population transport in thin films. The microscope achieves a temporal resolution down to 12 fs and simultaneously provides sub-10 nm spatial accuracy. We demonstrate the capabilities of the microscope by revealing an ultrafast excited-state exciton population transport of up to 32 nm in a thin film of pentacene and by tracking the carrier motion in p-doped silicon. The use of few-cycle optical excitation pulses enables impulsive stimulated Raman microspectroscopy, which is used for *in-situ* verification of the chemical identity in the $100 - 2000$ cm$^{-1}$ spectral window. Our methodology bridges the gap between optical microscopy and spectroscopy allowing for the study of ultrafast transport properties down to the nanometer length scale.

**TOC GRAPHICS**

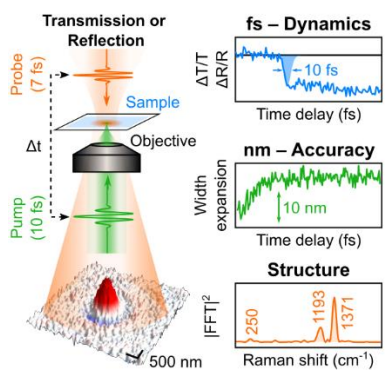

The performance of all optoelectronics materials is critically dependent on efficient carrier transport. Our understanding of the associated transport characteristics commonly relies on the interpretation of indirect measurements such as microwave conductivity, photoluminescence quenching, terahertz photoconductivity or time-of-flight techniques, which determine charge carrier mobilities or diffusion constants.[1–5] While these methods successfully describe transport in spatially-homogenous materials, they lack access to nanoscopic morphologies which can affect the deduced transport parameters. In particular, next-generation semiconductors such as hybrid metal-halide perovskites[6], organic semiconductors[7], quantum dot films[8,9] and monolayer 2D semiconductors[10] exhibit spatial heterogeneity on length scales as small as 10 nm, which need to be accessed to fully understand and optimize their charge carrier transport behavior in devices.

The photophysics of optoelectronic materials often involve sub-100 fs photodynamics, typically resolved using femtosecond transient absorption or reflection spectroscopy.[7,11,12] These ultrafast processes are implicated to be of relevance in spatial charge carrier (or exciton) transport, but the associated transport parameters are indirectly inferred from the obtained kinetics, which are intrinsically averaged over all morphologies within the probe spot (50 – 100 μm).[7,13,14] In contrast, spatially-localized measurement techniques such as transient optical microscopy currently achieve a spatial accuracy of σ ~ 10 – 50 nm, but are limited to time resolutions of ~100 – 300 fs, insufficient to access the earliest photodynamics.[15–18]

Here we overcome these limitations and explore the impact of ultrafast processes on the spatio-temporal carrier dynamics in optoelectronic materials by introducing a time-resolved transient absorption and reflection microscope capable of delivering sub-10 nm spatial accuracy with temporal resolutions down to 12 fs. In thin pentacene films, our technique resolves rapid spatial expansion occurring during the singlet fission process, which is completed within 200 fs after photoexcitation. In contrast, in p-doped silicon we observe charge carrier transport dynamics matching previously published reports even at the earliest time scales.[19,20] In addition, our high temporal resolution allows us to chemically characterize the sample region in a universal fashion by impulsive stimulated Raman scattering.

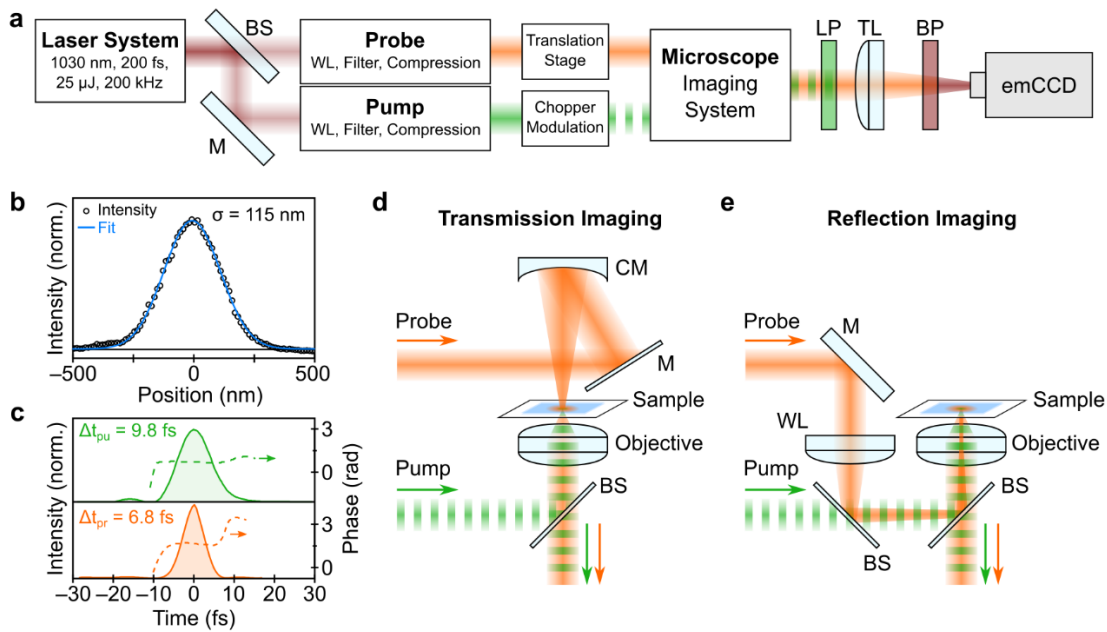

**Figure 1.** Setup and imaging channels employed in femtosecond transient absorption and reflection microscopy to monitor ultrafast charge carrier diffusion. **a**, Schematic of the optical components. A pump-probe scheme is integrated into a wide-field optical microscope with a high numerical aperture (NA = 1.1) objective and the transmitted or reflected probe is subsequently detected by an emCCD camera at a given probe wavelength. BS – beam splitter, LP – longpass filter, TL – tube lens, BP – bandpass filter, M – mirror. **b**, Spatial profile of the pump pulse obtained by monitoring the fluorescence intensity from a fluorescent bead scanned across the pump spot (black circles) and corresponding Gaussian fit (blue). **c**, Temporal characterization of the pump (green) and probe (orange) pulses at the sample. Probe pulses are only compressed for transmission experiments. **d**, Transmission geometry and **e**, Reflection geometry. CM – curved mirror, WL – wide-field lens.

Our experimental approach consists of an ultrafast pump-probe experiment coupled to a single-objective wide-field optical microscope (Figure 1a). Analogous to our previously reported implementation of wide-field femtosecond transient absorption microscopy,[12] we generate temporally compressed broadband pulses spanning wavelengths from 520 – 640 nm (pump) and 660 – 900 nm (probe). However, instead of using wide-field pump pulses we now focus the pump pulses tightly onto the sample by sending them through a high numerical aperture (NA = 1.1) oil-immersion objective, which results in a near-diffraction-limited Gaussian excitation profile ($\sigma$ = 115 nm, Figure 1b, supporting information, section 1.1). After the pump pulses have photoexcited the sample and generated a localized carrier population, we spatially resolve the photoinduced transient response by imaging time-delayed wide-field probe pulses onto an emCCD.

To achieve the highest temporal resolution, we employ a single-lens high-NA oil-immersion objective instead of the more conventional flat-field and chromatic aberration corrected multi-component objectives. This choice compromises the absolute imaging capability but only weakly affects a differential pump-probe experiment over a small field of view (see also supporting information, section 2), allowing us to compress our pump pulses to 9.8 fs with commercially available optical elements (Figure 1c, green line; supporting information, section 1.2). We remark that near-IR pulses (650 – 950 nm) can be temporally compressed to sub-10 fs for high-NA multi-component microscope objectives using advanced pulse-shaping methods.[21,22] Similar results for visible pulses (550 – 650 nm) are, however, lacking due to the significantly larger dispersion corrections required for this wavelength range.

The probe pulses in our setup are temporally compressed to 6.8 fs (Figure 1c, orange line) for the transmission channel, resulting in an effective time resolution of 11.9 fs (Figure 1d). In the reflection channel (Figure 1e), we use chirped probe pulses (~700 fs), which in combination with bandpass-filtered detection results in a temporal resolution of ~50 fs (see supporting information for details, section 1.2).[23] While it is straightforward to compress the probe pulses for our microscope objective by adding an additional set of third-order compensated chirped mirrors into the beam path, as illustrated for the pump pulse, we opted to demonstrate that chirped probe pulses can be used equivalently if a simpler implementation with marginally lower temporal resolution (~50 fs) is desired.

We incorporated two complementary probe channels to accommodate a broad range of samples irrespective of their optical properties. In the transmission channel, the probe is loosely focused onto the sample by a concave mirror (Figure 1d, Gaussian illumination area of σ ~ 8 μm), whereas in the reflection channel, the probe is focused into the back-aperture of the high-NA objective with an additional wide-field lens (Figure 1e, Gaussian illumination area of σ ~ 4 μm). Prior to detecting the probe, we insert narrow-bandpass filters into the detection path to gain access to wavelength-resolved spatio-temporal dynamics, while simultaneously removing optical artefacts associated with chromatic aberrations.

To demonstrate the capabilities of our microscope, we first imaged the spatio-temporal dynamics of a ~120 nm thin film of the singlet fission material pentacene (Pc) in transmission. Following photoexcitation to $S_1$, Pc undergoes singlet fission to form a correlated triplet-triplet ($^1$TT) state on a sub-100 fs time scale, which can be followed by monitoring the growth of the photoinduced absorption band at 790 nm. We remark that this wavelength range is free from other overlapping spectral components, allowing direct insight into the spatio-temporal characteristics of the singlet fission dynamics (supporting information, section 3.2).[24–27]

A typical point-spread function recorded at this wavelength 1 ps after photoexcitation exhibits the expected negative differential transmission signal at its center, accompanied by periodic intensity changes as we move radially away from the center (Figure 2a). Such a point-spread function is intrinsic to wide-field microscopy, also known as a far-field Frauenhofer diffraction pattern, which arises from the optical interference pathways of the probe light simultaneously collected over all space by the detector. In our experiment, the tightly focused pump pulses generate a Gaussian-shaped aperture by changing the complex refractive index of the material, through which the probe pulses are subsequently diffracted, yielding the characteristic near-isotropic Airy-disk point-spread function presented in Figure 2a. Closer inspection of our point-spread functions reveals a small degree of anisotropy in the intensity distribution of the first diffraction ring. We attribute this asymmetry to the reduced imaging capabilities provided by our microscope objective, causing spherical aberrations. This asymmetry does, however, not affect the central part of the point-spread function in our microscope, allowing us to confidently quantify time-resolved changes in its width (see supporting information, section 2 for a more detailed discussion).

The transient photodynamics of pentacene at the central pixel of the point-spread function feature an initial coherent spike at zero-time delay, which evolves into a fast rise of the $^1$TT population with a $\tau = 76 \pm 6$ fs time constant (Figure 2b, grey line). After the initial rise, the differential signal amplitude remains largely constant, in agreement with previous transient absorption studies.[24]

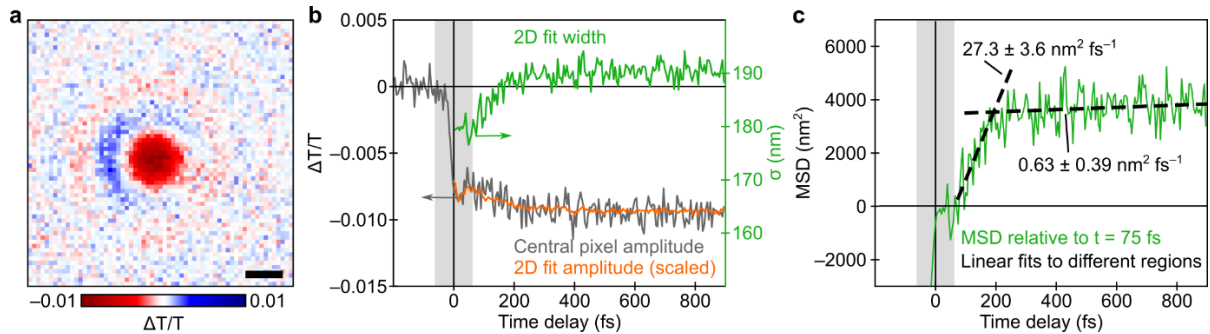

**Figure 2.** Femtosecond transient absorption microscopy of a thin pentacene film (~120 nm) measured in transmission. **a**, Differential transmission image at 1 ps after photoexcitation. Scale bar depicts 500 nm. **b**, Transient kinetics at the central pixel of the point-spread function (grey line, left axis) and retrieved amplitude of a two-dimensional (2D) Gaussian fit (orange line, left axis, scaled for clarity). The corresponding retrieved 2D Gaussian standard deviation (green line, right axis) is also shown. **c**, Mean-square displacement (MSD) curve relative to a time delay of 75 fs. The MSD is well described by two linear diffusion regimes with the indicated slopes (see main text for corresponding diffusion coefficient). Early time delays (<75 fs) were ignored due to coherent artefact contributions.

To isolate the underlying spatial profiles, we described each transient absorption image with an isotropic two-dimensional Gaussian function and extracted the corresponding time-dependent amplitudes (Figure 2b, orange line) and standard deviations (Figure 2b, green line), in line with previous analysis methods.[15,28] The retrieved amplitudes agree well with the central pixel amplitudes, but the standard deviations display an unexpected behavior characterized by an initial rise from ~180 nm at 75 fs to ~190 nm at 200 fs. This initial rise is followed by a significantly slower rise marginally above our signal to noise limit. We remark that this behavior is independent of fitting the point-spread function with an isotropic or anisotropic two-dimensional Gaussian function (supporting information, section 2).

Importantly, the spatial resolution of our microscope remains diffraction limited at the probe wavelength $\left(\sigma \sim \frac{\lambda}{2\,\text{NA}} \frac{1}{2\sqrt{2\ln2}} \sim 141 \text{ nm at } 790 \text{ nm}\right)$, but the spatial accuracy, i.e. how well we can distinguish the standard deviations of different transient point-spread functions is only dependent on the signal-to-noise ratio of the recorded images, as demonstrated previously.[15,28,29] Consequently, we can track changes to the profile of the imaged point-spread function with high confidence. By combining the retrieved standard deviation error from the two-dimensional Gaussian fit with the temporal standard deviation error we retrieve a spatial accuracy of ~2.9 nm for this measurement. We refer the reader to section 2 of the supporting information for more details (Figure S3). This spatial accuracy is well below the 10 nm standard deviation expansion observed during the first 200 fs for the pentacene films in Figure 2.

We can roughly estimate the underlying two-dimensional diffusion dynamics from the mean-square displacement (MSD) curves according to the (linear) diffusion Equation 1,

$$\text{MSD} = \sigma^2(t) - \sigma^2(t_0) = 2\,D(t - t_0) \tag{1}$$

where $D$ is the diffusion coefficient and $t$ the time delay (Figure 2c).[15,28,30] We note that we excluded early time delays in our analysis (Figure 2b,c, gray shaded area), such that $t_0 = 75$ fs, since the presence of coherent artefact contributions indicates additional signal pathways, such as two-photon absorption and cross-phase modulations, which can artificially reduce the retrieved width.[31,32] By analyzing the MSD of the previously identified standard deviation regimes, we obtain diffusion coefficients with values of $D_{\text{fast}} = 136.5 \pm 18.1$ cm$^2$ s$^{-1}$ and $D_{\text{slow}} = 3.2 \pm 2.0$ cm$^2$ s$^{-1}$. The initial diffusion coefficient is furthermore independent of excitation density in the linear absorption regime (Figure S5) eliminating carrier-recombination artefacts from our measurements. Based on the diffusion coefficient and retrieved singlet fission lifetime, we can estimate the effective carrier diffusion length for this fully temporally resolved ultrafast regime as $L_\text{fast} = \sqrt{D\tau} = 32.2 \pm 2.1$ nm.

To generalize this result, we repeated the same experiment by detecting different probe wavelengths (supporting information, section 4). Within a probe range from 710 – 790 nm, coinciding with the photoinduced $^1$TT absorption band in Pc (Figure S7), we retrieve identical fast diffusion coefficients within error margins. These results firmly rule out refractive index artefacts[33] or competing overlapping spectral features[16] as responsible factors for this fast diffusion regime. The extraordinarily high diffusion coefficients ($D_\text{fast}$) and associated carrier lengths ($L_\text{fast}$) thus point to a previously unknown transport regime during the earliest times following photoexcitation in Pc.

While a detailed interpretation of the above results is beyond the scope of this work, the established photophysics of Pc allows us to provide a first qualitative explanation of our results.[24,30,34,35] We speculate that the initially delocalized photoexcited $S_1$ wavefunction could undergo ultrafast spatial broadening ($D_\text{fast}$) during the formation of a more delocalized $^1$TT wavefunction within a coherent transport regime. This rapid initial expansion is followed by a slower population expansion over several picoseconds and the lower diffusion coefficient ($D_\text{slow}$) indicates an incoherent hopping mechanism which could be associated with the formation of an entangled but spatially separated $^1$T…T state.[36] Further theoretical work is, however, critical to elucidate the origins of both the fast and slow regimes of the observed spatio-temporal behavior.

In addition to our ability to visualize exciton population transport in Pc during singlet fission, the high temporal resolution of our setup allows us to also monitor the chemical fingerprint of the excitation spot via impulsive stimulated Raman scattering for *in-situ* verification of the chemical composition of the probed sample region.[37–39] To highlight this capability, we conducted the same experiment on Pc at higher excitation densities, to achieve an increased signal-to-noise ratio, and resolved pronounced oscillatory modulations on top of the electronic background signal (Figure 3a, ~25 fs period, see inset). To extract the associated impulsive Raman spectrum, we computed the Fourier power spectrum of the oscillatory modulations after removing the slowly-varying electronic background signal (Figure 3b). The retrieved spectrum is in excellent agreement with the ground-state Raman spectrum of Pc thin films and is dominated by modes at 1193 (~28 fs period) and 1371 cm$^{-1}$ (25 fs period).[40] The oscillatory components are also reproduced by applying a two-dimensional Gaussian fitting routine outlined above (Figure 3, orange line). We remark that the appearance of such coherent oscillations in the time-domain corroborates the high temporal resolution deduced from the pulse characterization (supporting information, section 1.2).

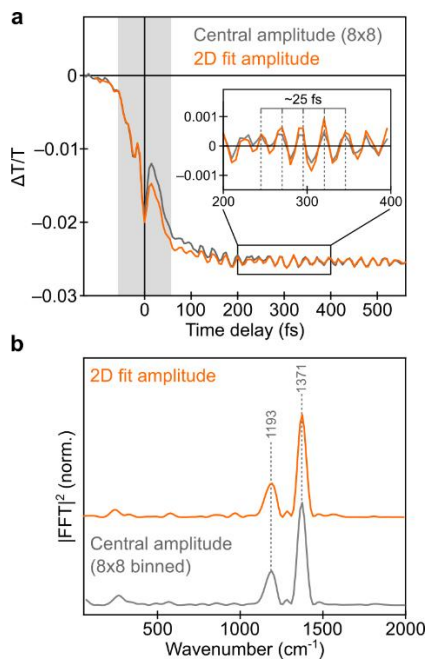

**Figure 3.** Vibrational imaging in femtosecond transient absorption microscopy demonstrated on pentacene in transmission. **a,** Detection of vibrational coherences superimposed on the electronic signal arising from impulsive stimulated Raman scattering. The oscillatory modulations are best visualized by either integrating the central image portion (8x8 pixels, grey line) or by carrying out a 2D Gaussian fit analysis (orange line). The signal is dominated by ~25 fs periods (inset). **b,** Fourier power spectrum after background removal of the electronic dynamics in **a** for the two analysis methods, displaying the dominant C-C and C=C stretching Raman modes of pentacene at 1193 and 1371 cm$^{-1}$.

To highlight how our approach can be applied for optically thick or non-transparent materials, we investigated the ultrafast spatio-temporal properties of p-doped Silicon (Si) using the reflection channel of our microscope. Si is a prototypical, well-characterized material commonly used in solar cells, which is non-transparent to visible light and therefore cannot be measured in transmission. Here, we recorded transient reflection images probed above the optical bandgap (740 nm). A typical point-spread function at a time delay of ~3 ps shows of a negative, isotropic differential reflectivity signal in the form of an Airy disk, in agreement with previous reports (Figure 4a).[33] The associated photodynamics exhibit an initial rise with a decay time constant of 207 ± 22 fs followed by a slower decay time of 11.29 ± 1.1 ps (Figure 4b, grey line).

Analogous to our analysis on Pc, we described each transient reflection image with a two-dimensional isotropic Gaussian function to analyze the spatial profiles. The retrieved time-dependent amplitudes match the central pixel amplitudes (Figure 4b, orange line), as previously observed for Pc. Interestingly, despite the two temporal regimes identified in the signal amplitude, the corresponding standard deviation only displays a single rise from ~190 nm at 100 fs to 205 nm at 3 ps (Figure 4b, green line), sufficient to be resolved with the spatial accuracy of ~2.7 nm obtained in this experiment (see supporting information, section 2). The corresponding diffusion coefficient in our time window can be estimated from the MSD using equation (1) and yields D = 8.6 ± 0.5 cm$^2$ s$^{-1}$, in agreement with previously reported values (Figure 4c).[19,33] Our results suggest that the diffusion characteristics in Si at the earliest times (~100 fs – 3 ps) after photoexcitation are equivalent to the pico- or nanosecond time scales that have been reported with alternative measurements techniques.[19,20,33]

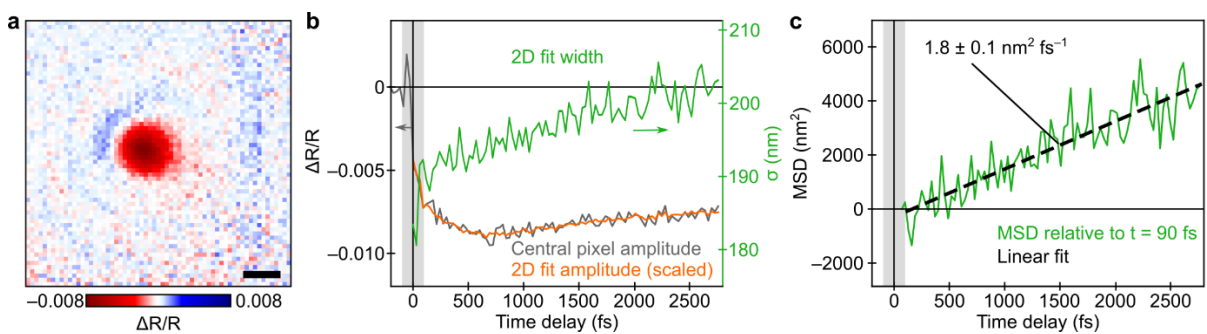

**Figure 4.** Femtosecond transient reflection microscopy of p-doped Silicon. **a**, Differential reflectance image at 2.8 ps after photoexcitation. Scale bar depicts 500 nm. **b**, Transient kinetics at the central pixel of the point-spread function (grey line, left axis) and retrieved amplitude of a two-dimensional (2D) Gaussian fit (orange line, left axis). The corresponding retrieved 2D Gaussian standard deviation (green line, right axis) is also shown. **c**, Mean-square displacement (MSD) curve relative to a time delay of 90 fs. The MSD is well described by a single linear diffusion regime with the indicated slopes (see main text for corresponding diffusion coefficient). Early time delays (<90 fs) were ignored due to coherent artefact contributions.

Apart from the improved temporal resolution, our transient absorption and reflection microscope differs significantly from previous implementations due to its wide-field imaging detection. Earlier studies[15–18] almost exclusively rely on confocal microscopy,[33] in which the transient image is reconstructed after scanning a tightly focused probe beam across the sample. The benefits and drawbacks of wide-field microscopy compared to confocal microscopy have been extensively reviewed in the literature and apply similarly for transient microscopes.[41] While we have demonstrated an ultrafast wide-field microscope due to its easier construction, our approach can readily improve the time resolution of a transient confocal microscope by simply replacing the relevant objective lenses with low-dispersion objectives and adding commercially available pulse compression optics (see Methods).

We emphasize that care has to be taken in the interpretation and comparison of the respective transient point-spread functions, which will appear as Airy-disks for a wide-field microscope (e.g. Figure 2a) instead of a two-dimensional Gaussian function for a confocal microscope.[15–18] Critically, while the underlying phase interference effects causing the Airy-disk point-spread functions in a wide-field microscope may seem deleterious at first, they in principle allow to monitor three-dimensional carrier motion on ultrafast time scales with our approach.[33,42]

Taken together, we have demonstrated ultrafast transient absorption and reflection microscopy which meets the time- and length scale requirements of optoelectronic processes by achieving an effective temporal resolution down to 12 fs in transmission and ~50 fs in reflection mode, with simultaneous sub-10 nm spatial accuracy. This spatio-temporal capability enabled us to uncover a previously unobserved ultrafast linear transport regime in Pc films indicative of a highly spatially delocalized $^1$TT wavefunction which corresponds to carrier transport lengths of up to 32 nm away from the initial excitation spot within only 200 fs after photoexcitation. In contrast, p-doped Si did not show any ultrafast spatial expansion but instead followed its reported long-time diffusive behavior. An inherent benefit of our approach is the ability to simultaneously monitor the chemical identity of the selected region via impulsive stimulated Raman scattering alongside the spatio-temporal dynamics. The combination of these functionalities provides comprehensive insight into charge carrier and exciton population transport in a wealth of materials with morphological and chemical specificity. Consequently, our approach can be readily applied to hetero-junctions and other inhomogeneous morphologies to elucidate interfacial population transport.

# EXPERIMENTAL METHODS

### Experimental setup.

Pulses are provided by a Yb:KGW amplifier (1030 nm, 200 fs, 200 kHz, 5 W, Pharos, Light Conversion) which seeded two whitelight continuum stages for probe and pump pulse generation. Probe pulses are produced in a 3 mm YAG crystal and spectrally filtered in a home-build 4F single prism filter to adjust the bandwidth to 650 – 900 nm. Subsequently, the probe pulses were compressed to 6.8 fs with third-order corrected chirped mirrors (DCM9, Venton) and by optimizing the prism insertion in the spectral filter. Probe pulses were only compressed for the transmission channel (see further explanation in supporting information, section 1). Consequently, they experienced a temporal chirp of ~700 fs in the spectral region from 650 – 900 nm in the reflection channel due to the additional dispersion caused by the wide-field and objective lens. Pump pulses were produced in a 3 mm Sapphire crystal for increased bandwidth, spectrally adjusted by a 650 nm shortpass filter (FESH650, Thorlabs) and subsequently compressed to ~10 fs by two sets of third-order compensated chirped mirrors (109811, Layertec) in combination with a wedge-prism pair, accounting for all optical components including the microscope objective. We additionally cleaned the spatial mode of the pump pulses by means of a 40 μm pinhole, which is imaged into the back-aperture of the objective to achieve near-diffraction limited focusing at the sample. To set up the time-dependent pump-probe sequence, we modulated the pump pulses at 15 Hz via a mechanical chopper (Thorlabs) and temporally delayed the probe via a computer-controlled closed-looped piezo translation stage (P-625.1CL, PhysikInstrumente).

An emCCD (Rolera Thunder, QImaging) detected the transmitted or reflected probe pulses via a high numerical aperture oil-immersion objective (NA = 1.1, 100× magnification) in combination with an F = 500 mm tube lens (AC254-500-B, Thorlabs) resulting in 55.5 nm/pixel. To avoid pump light detection, we inserted a 650 nm longpass filter (FELH650, Thorlabs) in front of the emCCD in addition to a narrow bandpass filter (FB790-10 for pentacene and FB740-10 for silicon, Thorlabs) as shown in Figure 1a.

An automatic focus control loop based on total internal reflection of a reference continuous wave laser (405 nm) was used to stabilize the focus position via an objective piezo (NP140, Newport) in pentacene.[43] The same approach could not be applied to silicon. Instead, we measured 8 different sample locations (experimental time 4 min/sample location) and averaged the results. During this time scale no significant focus drift was observed as verified by monitoring the autofocus performance during the measurements on pentacene.

The camera exposure time was set to 9 ms for pentacene and 12 ms for silicon at a camera frame rate of 30 Hz. For the experiments on pentacene shown in Figure 2 and 3, the photoexcited carrier densities were $2.9 \cdot 10^{18}$ and $9.01 \cdot 10^{18}$ cm$^{-3}$ for Pc,[44] respectively, and $1.67 \cdot 10^{18}$ cm$^{-3}$ for silicon,[20] as calculated from the peak energy of the pump pulse.[33] Both experiments were conducted by adjusting the polarization of the pump and probe beams to magic angle.

**Data analysis.**

Normalized differential images were calculated from the raw images and described for each time delay with an isotropic two-dimensional Gaussian function according to:

$$G(x,y) = A \exp\left(-\frac{(x-x_0)^2+(y-y_0)^2}{2\sigma^2}\right) + \text{offset}$$

where $A$ encodes the signal amplitudes, $x_0$ and $y_0$ the position and $\sigma$ is the standard deviation. Initial guesses for the position were taken from the center of mass position of each image, the amplitude was taken from the corresponding pixel amplitude and we set $\sigma = 100$ nm and offset = 0.

We note that this approach does not describe the diffraction rings present in the Airy-disk. This results in a systematic overestimation of the fitted signal amplitude and we therefore scaled the reported amplitudes in Figure 2 and 4 to emphasize the correct reproduction of the transient kinetics. Importantly, our analysis focusses on evaluating time-dependent differences to the retrieved fit parameters. Especially for the standard deviation parameter, any error in describing the full point-spread function will translate into a small systematic deviation from the actual value, which will be inconsequential for assessing temporal changes.

Population kinetics were modelled with a sum of two exponentially decaying Gaussian functions, and stated error bars in the retrieved lifetimes correspond to the standard deviation error.

ASSOCIATED CONTENT

**Supporting Information**.

Supporting information. further information regarding the temporal and spatial pulse characterization, illustration of the general fit performance and accuracy, sample preparation and characterization details, discussion on points-spread function.

**Additional Information.**

The data supporting this publication is available free of charge at [*URL to be added in proof*]

AUTHOR INFORMATION

**Notes**

The authors declare no competing financial interests.

**ORCID**

Christoph Schnedermann: 0000-0002-2841-8586

Jooyoung Sung: 0000-0003-2573-641

Raj Pandya: 0000-0003-1108-9322

Sachin Dev Verma: 0000-0002-6312-9333

Nicolas Gauriot : 0000-0001-7725-7208

Hope Bretscher: 0000-0001-6551-4721

Philipp Kukura: 0000-0003-0136-7704

Akshay Rao: 0000-0003-4261-0766

ACKNOWLEDGMENT

We acknowledge financial support from the EPSRC and the Winton Program for the Physics of Sustainability. This project has received funding from the European Research Council (ERC) under the European Union's Horizon 2020 research and innovation program (grant agreement No 758826). C.S. acknowledges financial support by the Royal Commission of the Exhibition of 1851.